\documentclass[pdflatex,sn-mathphys-num]{sn-jnl}

\usepackage{graphicx}%
\usepackage{multirow}%
\usepackage{amsmath,amssymb,amsfonts}%
\usepackage{amsthm}%
\usepackage{mathrsfs}%
\usepackage[title]{appendix}%
\usepackage{xcolor}%
\usepackage{textcomp}%
\usepackage{manyfoot}%
\usepackage{booktabs}%
\usepackage{algorithm}%
\usepackage{algorithmicx}%
\usepackage{algpseudocode}%
\usepackage{listings}%
\usepackage[switch]{lineno}

\theoremstyle{thmstyleone}%
\theoremstyle{thmstyletwo}%

\theoremstyle{thmstylethree}%

\begin{document}


\title[Monolithic high density integrated photonics on bulk lithium niobate]{Monolithic high density integrated photonics on bulk lithium niobate}

\author*[1]{\fnm{Zizheng} \sur{Li}}\email{z.l.li-1@tudelft.nl}

\author[2]{\fnm{Harmen} \sur{Smedes}}

\author[1]{\fnm{Bruno} \sur{Lopez-Rodriguez}}

\author[1]{\fnm{Thomas} \sur{Scholte}}

\author[2]{\fnm{Simon} \sur{Gröblacher}}

\author*[1]{\fnm{Iman} \sur{Esmaeil Zadeh}}\email{i.esmaeilzadeh@tudelft.nl}

\affil[1]{\orgdiv{Department of Imaging Physics}, \orgname{Faculty of Applied Sciences, Delft University of Technology}, \city{Delft}, \postcode{2628 CJ}, \country{The Netherlands}}

\affil[2]{\orgdiv{Department of Quantum Nanoscience}, \orgname{Faculty of Applied Sciences, Delft University of Technology}, \orgaddress{\city{Delft}, \postcode{2628 CJ}, \country{The Netherlands}}}


\abstract{
Functional electro-optic (EO) tunable photonic integrated circuits are crucial to next-generation information processing and advanced computing, where ferroelectric materials such as lithium niobate (LiNbO$_3$) provide outstanding optical properties and versatile tuning mechanisms.  
However, beyond achieving high-performance devices, practical deployment of LiNbO$_3$ photonic integrated circuits requires rapid and cost-effective scale-up, which remains challenging because of the fabrication complexity and high costs of ion-sliced thin-film lithium niobate and wafer bonding processes. 
To address this gap, we demonstrate a fully monolithic photonic platform, amorphous silicon carbide (a-SiC) on bulk LiNbO$_3$ crystal substrate, enabling scalable, CMOS-compatible, low-cost, and high-density photonic integrated circuits without relying on thin-film LiNbO$_3$. This integration approach breaks the long-standing cost–performance–scalability tradeoff, and intrinsically eliminates the need for sophisticated fabrication processes, including ion slicing, wafer bonding, and LiNbO$_3$ etching. It realizes high-density, low-loss, and record high EO tuning efficiency of $V_{\pi}L=$ 2.87 V$\cdot$cm on bulk LiNbO$_3$ crystal substrate. Furthermore, leveraging slow-light effect near photonic crystal bandgap, an 8.5-fold EO tuning efficiency enhancement is achieved, highlighting the platform's capability in confinement control and dispersion engineering. 
}

\maketitle

\section{Introduction}\label{sec1}

By miniaturizing the conventional bulk optical elements and integrating them on a photonic chip, photonic integrated circuits (PICs) provide a compact and scalable approach to on-chip light manipulation, bolstering a wide range of technologies, including telecommunications, precision metrology, and next-generation computing. For PICs, beyond device-level performance metrics such as low propagation loss,  strong nonlinear effects, and efficient tuning mechanisms, the practical deployment of PICs in emerging technologies is ultimately determined by scalability, integration density, and fabrication cost.

Despite the prevalence of silicon (Si) and silicon nitride (SiN) PICs, which benefit from mature complementary metal oxide semiconductor (CMOS) fabrication processes, their performances are fundamentally constrained by two-photon absorptions in Si and the lack of intrinsic tuning mechanisms in SiN.\cite{tsang2002optical,xiang2022silicon,alexander2018nanophotonic} Among emerging PIC material platforms with exceptional properties, lithium niobate (LiNbO$_3$ or LN) is widely recognized as one of the most promising materials for next-generation integrated photonics. It possesses distinct combination of properties including wide transparency window, low material loss, strong second-order nonlinearity, diverse tunability mechanisms such as electro-optic, acousto-optic and piezoelectric effects, and compatibility with rare-earth doping.\cite{zhu2021integrated,boes2023lithium,honardoost2020rejuvenating,guarino2007electro,wang2018nanophotonic,hu2025integrated,zhang2017monolithic,li2026ultrabroadband} These properties have established LN as a leading platform for high-speed modulators and switches, making it particularly attractive for fast and non-thermal reconfigurable PICs in applications such as photonic computing, telecommunications, and quantum information processing. 

Although lithium niobate photonics has advanced considerably, the fabrication of integrated LiNbO$_3$ devices remains a formidable challenge. Methods such as ion diffusion and proton exchange have been applied to form optical waveguides on bulk LiNbO$_3$ substrates for optical modulators. However, the inherently low index contrast between the guiding and surrounding materials in these approaches results in weak optical confinement, necessitating large minimum bending radii and limiting EO tuning efficiency, hampering their way towards high density integration and large scale photonic integrated circuits (PICs).\cite{zhu2021integrated,wooten2000review,januar1993characteristics} 
Smart-Cut technology has enabled the transfer of thin-film LiNbO$_3$ (TFLN) layers onto Si-based substrates, providing strong optical confinement and allowing for compact devices.\cite{ramadan2000electro,guarino2007electro,rabiei2004optical,poberaj2009ion} This technology has transformed the fabrication of LiNbO$_3$ photonic integrated circuits, and has attracted rapidly growing research interest worldwide. Nonetheless, Smart-Cut technology based TFLN fabrication relies on multiple critical processes including ion-implantation, ion-slicing and wafer-bonding, and typically necessitates delicate pre- and post-bonding treatments, such as chemical-mechanical polishing (CMP) and thermal annealing, to ensure device feasibility and mitigate implantation-induced crystal damage. These intricate procedures impose outstanding challenges on the subsequent fabrication steps, for example a limited thermal budget, residual crystal damages and defects, and unwanted strain introduced during bonding.\cite{mookherjea2023thin,shams2022reduced,jia2021ion,huang2014helium,poberaj2012lithium,dolabella2023correlated,takigawa2019residual,prabhakar2021stress} It also substantially increases the fabrication cost and process complexity, posing a major obstacle to the rapid scale-up of TFLN PIC production and the industrial-scale deployment. Furthermore, the etching of LiNbO$_3$ has been recognized as a persistent challenge and is considered incompatible with standard CMOS foundry processes due to the risk of lithium contamination.\cite{jia2021ion,kumar2024optimization,li2023high,kaufmann2023redeposition} Thus, alternative approaches that bypass these fabrication complexities while harnessing LiNbO$_3$'s exceptional material properties are highly valuable.

To avoid direct etching of LiNbO$_3$ and thereby reduce fabrication complexity, hybrid and heterogeneous integration strategies have been demonstrated using various materials for rib-loading waveguides.\cite{churaev2023heterogeneously,mao2022heterogeneous,huang2021high,yu2019photonic,jin2019mid,nakanishi2010high,li2025heterogeneous,zhang2021integrated,krishna2024hybrid} 
However, all of these demonstrations still rely on the TFLN substrate, predominantly prepared through ion slicing and wafer-bonding processes, which are not CMOS-compatible and fundamentally limit the rapid, cost-effective scale-up of LiNbO$_3$ photonics for expanding application demands. 
As a result, fully monolithic photonic integration directly on bulk LiNbO$_3$ crystals, while simultaneously enabling dense integration and low driving-voltage EO operation, remains a highly desired yet elusive goal.

An early effort toward fully monolithic LiNbO$_3$ photonic integration, based on chemical-vapor-deposited hydrogenated amorphous silicon (a-Si:H) on LN, was demonstrated in 2014\cite{cao2014hybrid}. However, its further development can be limited by the high optical loss of a-Si:H at telecom wavelengths.\cite{bristow2007two,tsang2002optical,fan2015optical} 
Amorphous silicon carbide (a-SiC) as an emerging photonic integration material, distinguished by low-loss, strong third-order nonlinearity, and broad compatibility, offers a promising route to addressing this gap in scalable LiNbO$_3$ photonic integration.\cite{xing2019cmos, lopez2023high, li2025heterogeneous, li2025heterogeneous_sin}

In this work, we introduce a fully monolithic a-SiC/LN photonic integration platform based entirely on CMOS-compatible fabrication processes, with all a-SiC deposition and patterning steps performed below 300$^{\circ}$C while leaving the underlying LiNbO$_3$ unmodified. This approach features a notably simple and broadly compatible route towards scalable LiNbO$_3$ PICs. On the one hand, the fabrication of this platform bypasses the rigorous pre- and post-processing associated with smart-cut and wafer bonding in TFLN-based integration approaches. On the other hand, it eliminates the need for direct etching of lithium niobate, while enables low-loss, high-density, and efficiently tunable LiNbO$_3$ photonic integration using a-SiC loading ribs. In the a-SiC/LN platform, on-chip light guiding, routing, and thermo-optic (TO) tuning are enabled by photonic waveguides patterned in the high-refractive-index a-SiC layer, while a controlled fraction of the optical mode remains confined in the bulk LN substrate, allowing access to the intrinsic material properties of LiNbO$_3$. Ring resonators are fabricated to quantify waveguide losses, thermo-optic and electro-optic (EO) tunability. Dispersion engineered photonic crystal (PhC) devices further demonstrate the platform's ability to precisely control optical confinement and photonic dispersion on unpatterned bulk LiNbO$_3$, despite the absence of the high index-contrast normally afforded by TFLN platforms. Through the PhC-induced slow-light effect, an 8.5-fold enhancement in EO tuning efficiency is achieved, underscoring a substantial enhancement of the electro-optic light–matter interaction.

Beyond LiNbO$_3$, the broad compatibility of this monolithic a-SiC rib-loading integration strategy allows for its extension to other optical crystal materials, including barium titanate (BTO), strontium titanate (STO), lithium tantalate (LiTiO$_3$), aluminium nitride (AlN), tantalum pentoxide (Ta$_2$O$_5$), lead zirconate titanate (PZT), and other functional optical crystals. This fully monolithic integration approach therefore provides a broadly applicable and scalable route for photonic integration on diverse crystalline substrates, opening opportunities in optical communications, nonlinear photonics, and quantum photonics.

\section{Results}\label{secresults}

\subsection*{Fabrication of amorphous silicon carbide on bulk lithium niobate platform}

Fig. 1\textbf{a} represents the fabrication flow of the a-SiC/LN platform. The fabrication begins with the chemical vapor deposition of amorphous silicon carbide thin-films onto commercially available 500 $\mu$m X-cut bulk lithium niobate substrates. On the deposited a-SiC layer, photonic structures are patterned by electron-beam (E-beam) lithography and transferred via fluorine based reactive ion etching. This lithography-etching process can be repeated, when required for partially etched photonic crystal structures. Electrodes are then defined by metal liftoff process, and the a-SiC waveguides are covered by SiO$_2$ as optical cladding. The a-SiC/LiNbO$_3$ fabrication process remains fully CMOS compatible by avoiding any direct processing of lithium niobate and maintaining the process temperature below 300 $^{\circ}$C. 

\begin{figure}[h]
\centering
\includegraphics[width=0.9\textwidth]{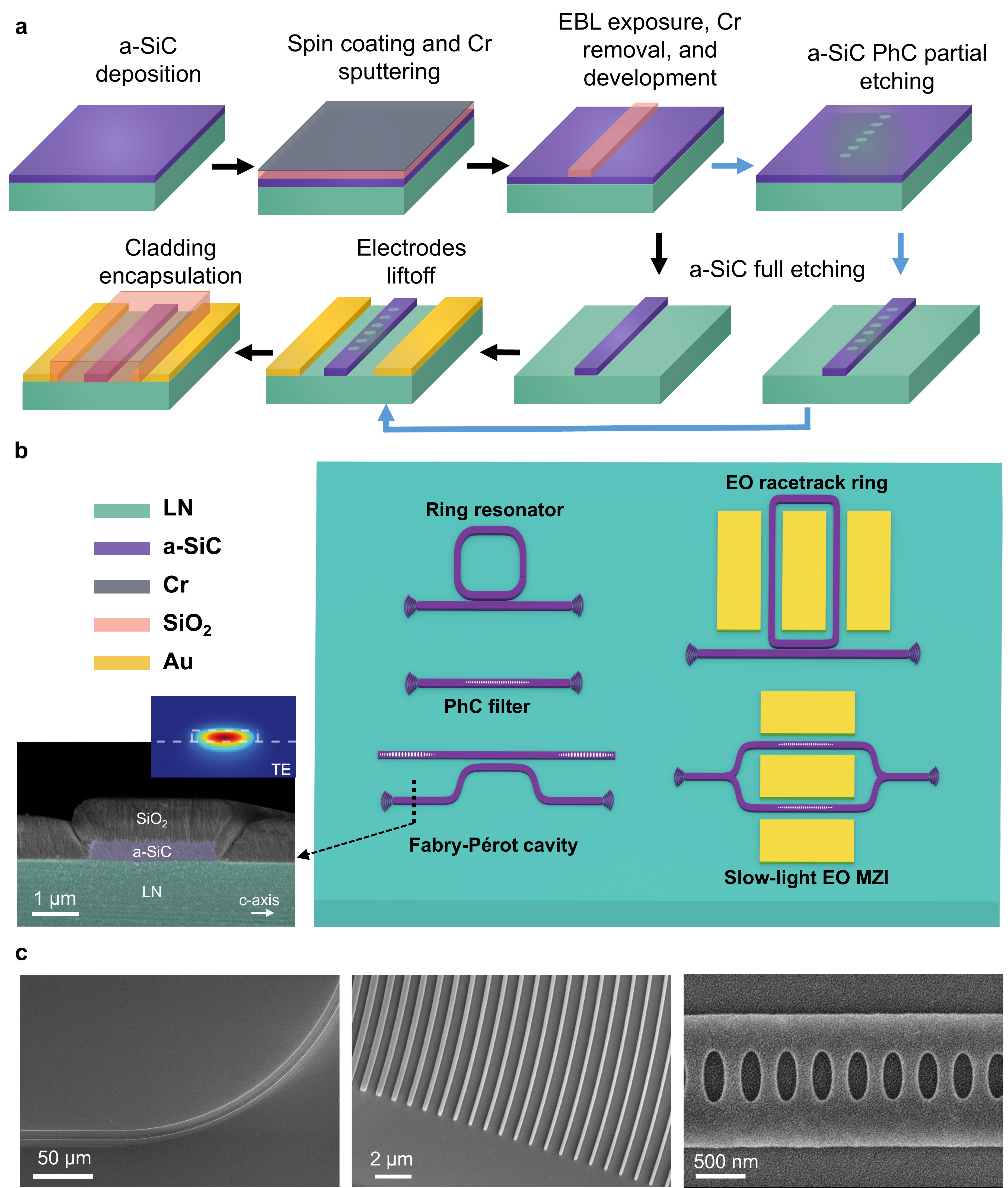}
\caption{\textbf{Monolithic integration of a-SiC on bulk LN photonics.} \textbf{a,} Fabrication processes. \textbf{b,} Schematic representation of a-SiC/LN functional devices: Passive microring resonator (top left), tunable racetrack resonator (top right), PhC filter (mid left), PhC Fabry-P\'{e}rot cavity (bottom left), and PhC slow-light MZI (bottom right). The inset on the left is an SEM image showing the cross-section of a-SiC/LN waveguide, with a simulated profile of the fundamental TE mode. \textbf{c,} Close-up SEM images of a waveguide bend, grating coupler, and PhC holes.}\label{fig1}
\end{figure}

The bottom-left part of Fig. 1\textbf{b} shows an SEM image of the cross-section of the a-SiC rib-loading waveguide on bulk LN substrate, and the inset depicts the electric-field distribution of the fundamental TE mode as the mode of interest, simulated using a finite-difference eigenmode solver (Lumerical MODE Solutions). With cross-sectional dimensions of 2.1 $\mu$m in width and 350 nm in thickness, the a-SiC rib-loading waveguide supports only the orthogonal fundamental modes, where the mode energy is distributed both in a-SiC and LiNbO$_3$. 

The schematic on the right side of Fig. 1\textbf{b} illustrates the designed a-SiC/LN photonic devices, including passive microring resonators and tunable racetrack resonators. To highlight a-SiC/LN platform's capabilities in dense-integration and dispersion engineering, devices based on photonic crystal (PhC) structures are designed and studied, including PhC filters and PhC Fabry-P\'{e}rot cavities. Finally, active PhC-based EO Mach–Zehnder interferometers (MZIs) are investigated to demonstrate slow-light enhanced electro-optic tuning on the a-SiC/LN platform. Fig. 1\textbf{c} shows the scanning electron microscope (SEM) images of the fabricated devices, featuring smooth waveguide bends, grating lines with a feature size of 60 nm, and consistent periodical partially-etched elliptical photonic crystal holes. Note that all SEM images presented in this work were taken from samples coated with a 10-nm-thick sputtered gold layer, in order to minimize the charging effect during SEM imaging on insulating bulk LiNbO$_3$ substrates. Therefore, the surface grains visible in the images most likely originate from this conductive coating rather than from the underlying surfaces.

\subsection*{Design and characterization of a-SiC/LN platform}

Propagation loss and device tunability are two key metrics of a photonic integration platform. We designed and fabricated optical ring resonators on the proposed a-SiC/LN platform to reveal its capabilities in low-loss propagation, efficient thermo-optic and electro-optic tuning.  

\begin{figure}[!htbp]
\centering
\includegraphics[width=0.85\textwidth]{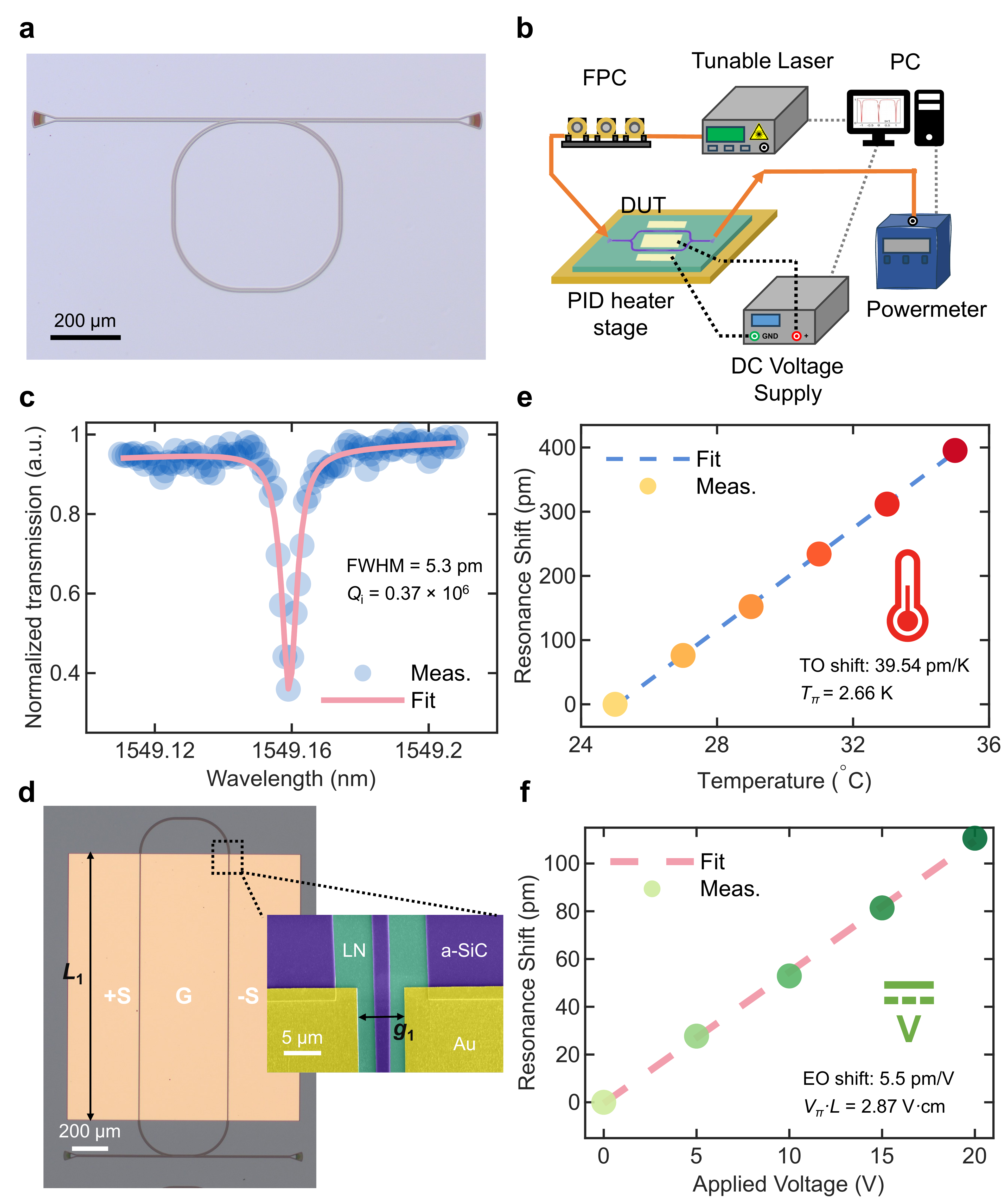}
\caption{\textbf{Optical property and tunability of a-SiC/LN platform.} \textbf{a,} Optical microscopy image of a fabricated a-SiC/LN microring resonator. \textbf{b,} Schematic representing the experimental setup used to characterize the a-SiC/LN devices. \textbf{c,} Measured resonance dip of the microring resonator at around 1549.16 nm, and the corresponding Lorentzian fit. \textbf{d,} Optical microscope image of a fabricated a-SiC/LN racetrack ring resonator, and a false colored SEM image showing a close-up view of the a-SiC waveguide between the Au electrodes. The electrodes are separated by a gap of $g_\text{1}=3.6$ $\mu$m, with the same length of $L_\text{1}=0.15$ cm. \textbf{e,} Thermo-optical response measured on the a-SiC/LN racetrack ring resonator, together with the linear fitting. \textbf{f,} Electro-optical response measured on the a-SiC/LN racetrack ring resonator, together with the linear fitting. } 
\label{fig2}
\end{figure}

To accommodate different application requirements, first, the dimensions of the a-SiC rib-loading waveguides are systematically analyzed. With the waveguide thickness fixed at 350 nm, three waveguide widths ($w=$2.6 $\mu$m, 2.1 $\mu$m, 1.35 $\mu$m) are investigated for low-loss propagation, high-efficiency tuning, and confinement engineering. The three geometric schemes are simulated and compared in terms of field distribution and bending loss, as demonstrated in Table. S1 in Supplementary Materials. All three geometric schemes ensure single-mode propagation, with only the orthogonal fundamental modes supported, and allow adiabatic transition between each other through linear waveguide tapering. For on-chip photonic routing and high-density integration, wide waveguide width (2.6 $\mu$m) is selected to minimize the impact of sidewall roughness arising from reactive ion etching, hence enabling low propagation losses. Bending losses for different rib-loading waveguide widths are simulated and compared in Fig. S1. The results indicate a negligible bending loss ($<1\times10^{-4}$ dB per bend) for $R\geq90$ $\mu$m when $w=$2.6 $\mu$m. For electro-optic tuning and modulation, a waveguide width of 2.1 $\mu$m is chosen for reduced electrodes gap distance, thereby enhanced electric-field intensity. The metal absorption losses coming from the gold (Au) electrodes for electro-optic tuning are simulated and presented in Fig. S2. In contrast, confinement engineering incorporating shallow-etched photonic crystal structures requires a narrow waveguide width of 1.35 $\mu$m.

Experimentally, we validate low-loss propagation on the a-SiC/LN platform on a fabricated microring resonators with 200 $\mu$m bending radius and 2.6 $\mu$m waveguide width (Fig. 2\textbf{a}). In fabrication, after a-SiC etching the microring resonators were encapsulated with a 1.5 $\mu$m thick plasma enhanced chemical vapor deposited (PECVD) SiO$_2$ layer as optical cladding. The microring resonators are characterized through grating couplers (see Supplementary Materials for details) on both sides as input and output interfaces, using the optical setup depicted in Fig. 2\textbf{b}. In Fig. 2\textbf{c}, a resonance dip of the microring resonator exhibits a full-width half-maximum (FWHM) of 5.3 pm, corresponding to a loaded quality factor of 292,051, and an intrinsic quality factor of $Q_\text{i}=$365,964, indicating a waveguide propagation loss of 1.3 dB/cm for the a-SiC/LN platform.

While electro-optic tunability has been a major motivation for using LiNbO$_3$, many applications also require quasi-static tuning to set or stabilize device operating points. Under such circumstances, thermo-optic tuning remains the preferred approach. However, LiNbO$_3$'s relatively low thermo-optical coefficient (TOC) necessitates large local temperature change to induce a $\pi$ phase shift, thereby posing challenges for on-chip thermal dissipation control.\cite{shim2026generalized} On the a-SiC/LN platform, efficient TO tuning is enabled by the large TOC of a-SiC, validated by a TO characterization on the racetrack ring resonator (Fig. 2\textbf{d}). The sample was mounted on a proportional-integral-derivative (PID) controlled heater stage, where the microring resonance shift was measured while changing the temperature from 25 $^{\circ}$C to 35 $^{\circ}$C. The resonance wavelength shifts linearly with temperature, exhibiting a thermo-optic tuning slope of 39.54 pm/K (Fig. 2 \textbf{e}), corresponding to a $\pi$-equivalent phase tuning temperature of $T_\pi=$ 2.66 K. This high thermal tuning efficiency enables precise operating point control while requiring only a small local temperature change.  

To evaluate the EO tuning efficiency of a-SiC/LN platform, the Pockels effect is measured on the same racetrack ring resonator illustrated in Fig. 2d. The racetrack ring resonator has two straight waveguide segments sandwiched between Au electrodes, with a length of $L_1=$ 0.15 cm. Depicted in the magnified SEM view in the inset, the Au electrodes are spaced by a gap of $g_1=$ 3.6 $\mu$m, while the waveguide width is 2.1 $\mu$m. Electro-optic tuning of the racetrack ring is driven by Au electrodes configured in a positive-signal, ground, negative-signal (+S,G,-S) arrangement. As a result, the EO-driven resonance shift is observed to be 5.5 pm/V, corresponding to a half-wave voltage-length product of $V_{\pi}L=$ 2.87 V$\cdot$cm, given the free spectral range (FSR) of the ring of 210 pm. To the best of our knowledge, it is the highest EO tuning efficiency achieved in photonic integrated circuits on bulk LiNbO$_3$. Nevertheless, we show in the following sections that slow-light effect can enhance the electro-optic tuning efficiency a factor of 8.5, leveraging a dispersion engineered one-dimensional photonic crystal structure on the a-SiC/LN platform.

\subsection*{Photonic crystal structures on a-SiC/LN platform}

\begin{figure}[!htbp]
\centering
\includegraphics[width=0.885\textwidth]{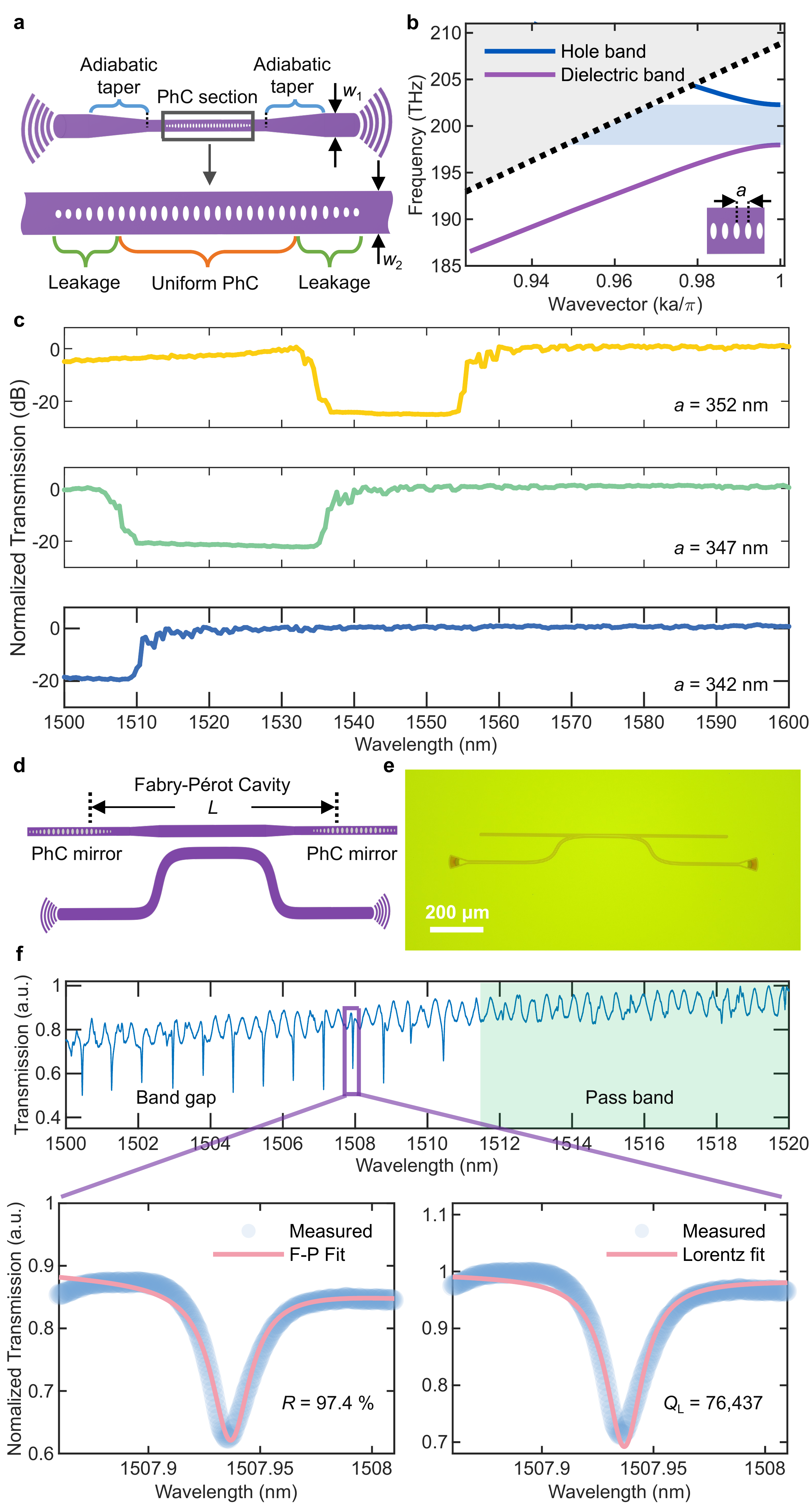}
\caption{\textbf{Photonic crystal structures on a-SiC/LN platform.} \textbf{a,} Schematic of an a-SiC/LN PhC waveguide filter, with a magnified view of the photonic crystal structure region. The photonic crystal structure comprises a central section of uniform periodic holes (PhC) and size-tapered leakage holes (Leakage) on both sides. The input-and-output waveguides are adiabatically tapered from width $w_1=2.1$ $\mu$m down to $w_2=1.35$ $\mu$m. \textbf{b,} Simulated bandstructure of the PhC structure. \textbf{c,} Measured transmission spectra of three a-SiC/LN PhC waveguide filters with lattice constants of $a=342$ nm, $a=347$ nm, $a=352$ nm, respectively. \textbf{d,} Schematic of an a-SiC/LN Fabry-P\'{e}rot cavity constructed using two PhC mirrors, with a bus waveguide evanescently coupled to inject and extract light. \textbf{e,} Optical microscopic image of a fabricated a-SiC/LN Fabry-P\'{e}rot cavity. \textbf{f,} Top: Measured transmission spectrum of the Fabry-P\'{e}rot cavity, showing resonances only inside the bandgap and absent in the passband. Bottom: Magnified view of a resonance dip that fitted to F-P function and Lorentz function (Methods). A PhC mirror reflectivity $R=97.4\%$ and a loaded quality factor $Q_\text{L}=76,437$ are obtained.}
\label{fig3}
\end{figure}

In the absence of bottom index contrast, integrated photonics on bulk LiNbO$_3$ lacks the strong vertical optical confinement typically provided by the substrate in LNOI and other conventional photonic platforms. Therefore, realizing strong optical confinement and dense photonic integration on bulk LiNbO$_3$ is highly significant, meanwhile challenging. To reveal the capability of the monolithic a-SiC/LiNbO$_3$ integration platform for optical confinement control, partially etched a-SiC photonic crystal (PhC) structures are employed to realize on-chip optical filters and reflectors.\cite{huang2016ultra,yu2019photonic_Kerr} 

As shown in Fig. 3\textbf{a}, the a-SiC/LN optical filter device consists of a PhC section, connected to straight waveguides on both ends, which are adiabatically tapered from width $w_2=$2.1 $\mu$m to $w_1=$1.35 $\mu$m and terminated by grating couplers for fiber-chip interfacing. With a waveguide width of $w_2=$2.1 $\mu$m, achieving sufficiently strong effective-index modulation to open a pronounced photonic bandgap would require large elliptical etched holes, resulting in prohibitively small feature sizes in fabrication. We therefore taper the waveguide to a narrower width, in order to enhance the relative change of effective index without compromising the fabrication fidelity. The PhC section comprises partially-etched elliptical holes in the a-SiC layer, including a central uniform PhC section, and two leakage sections on both sides, in which the major axis length of the ellipses gradually varied (details in Supplementary Materials) to suppress the radiation loss during the transition between the regular waveguide and the PhC structure. In Fig. 3\textbf{b}, with a lattice constant of $a=$ 337 nm, the band structure of the uniform PhC section is simulated (COMSOL Multiphysics), assuming a 1D lattice with infinitely periodic elliptical holes. Below the light cone (gray shaded area), two guided bands are present, corresponding to the etched-hole mode and dielectric mode, respectively. A photonic bandgap opens between 198 THz and 202 THz, corresponding to a wavelength range of 1482 nm to 1514 nm. Experimentally, a-SiC/LN filters with different lattice constants around 337 nm are fabricated and characterized (Fig. 3\textbf{c}). 
 The measured bandgap width (32 nm) agrees well with the simulated value, while the simulated bandgap position for $a=$ 337 nm overlaps with the measured spectrum of the fabricated a-SiC/LN filter with a nominal lattice constant of $a=$ 342 nm. This mismatch is attributed to a 5 nm shrinkage in the lattice constant during fabrication. The a-SiC/LN filters exhibit an extinction ratio of approximately 20 dB and clear band edges. These results confirm that solely introducing partially etched a-SiC PhC structures can provide effective dispersion engineering on bulk LiNbO$_3$. At shorter wavelengths (1500 - 1530 nm), the measured spectra show an upward trend in the transmission baseline, arising from the wavelength-dependent grating coupler response in this spectral range, as detailed in Fig. S4 in Supplementary Materials. 

To further quantify the dispersion-engineering and confinement-control capabilities of the a-SiC/LN platform, we design a Fabry-P\'{e}rot cavity by connecting two PhC mirrors (leakage-uniform-leakage structure) with a straight waveguide evanescently coupled to a bus waveguide, as depicted in Fig. 3\textbf{d}. A lattice constant of 342 nm is employed for the PhC mirrors. Similar to the PhC filter discussed in Fig. 3\textbf{a}, in the Fabry-P\'{e}rot cavity the straight cavity waveguide is tapered to a width of $w_2=$1.35 $\mu$m at both ends before the PhC mirrors. The cavity length (the distance between two PhC mirrors) is approximately \textit{L} = 637 $\mu$m. The transmission spectrum of the fabricated device (Fig. 3\textbf{e}) is measured over the wavelength range of interest from 1500 nm to 1520 nm, is shown in the upper panel of Fig. 3\textbf{f}. The PhC mirrors act as reflectors only within the PhC bandgap and remain transmissive in the PhC passband, consequently, Fabry-P\'{e}rot resonance dips are only observed within the PhC bandgap. 

A representative cavity resonance at 1507.9 nm is presented in the lower panels of Fig. 3\textbf{f}. By fitting the resonance dip using a Fabry-P\'{e}rot cavity transmission model (F-P Fits, described in Eq.(1) in Methods) and a Lorentzian function, we extract a mirror reflectivity of 97.4$\%$ and a loaded quality factor of 76,437, respectively. The measured quality factor is comparable to the highest values reported for suspended amorphous silicon carbide photonic crystal cavities\cite{lee2015high}, despite the reduced vertical refractive index contrast imposed by directly integrating the a-SiC structures on bulk LiNbO$_3$. These results highlight the capability of the a-SiC/LiNbO$_3$ platform to achieve strong optical confinement through PhC structures defined in a single a-SiC layer.

\subsection*{Slow-light enhanced efficient electro-optical tuning}

\begin{figure}[!htbp]
\centering
\includegraphics[width=0.885\textwidth]{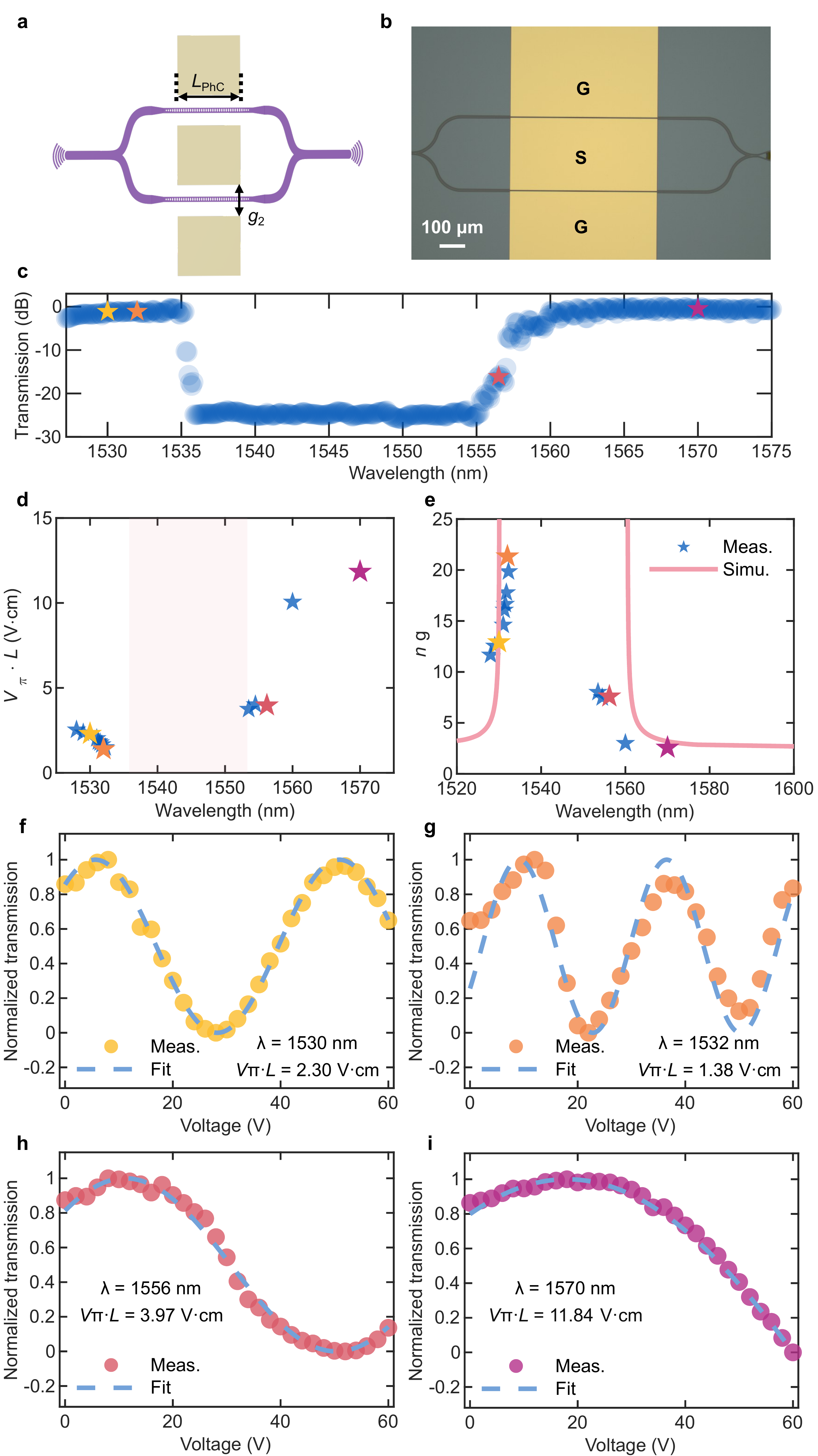}
\caption{\textbf{Slow-light enhanced electro-optic tuning.} \textbf{a,} Schematic of an a-SiC/LN PhC slow-light MZI, with an electrode length $L_\text{2}=0.06$ cm and a gap $g_\text{2}=4.0$ $\mu$m. \textbf{b,} Optical microscopy image of the fabricated slow-light MZI. \textbf{c,} Measured transmission spectrum of the slow-light MZI when zero voltage is applied. \textbf{d,} Measured half-wave voltage-length products of the slow-light MZI. \textbf{e,} Corresponding effective group indices extracted from the EO measurements, together with the estimation given by simulations. Normalized transmission of the slow-light MZI when sweeping voltages at \textbf{f,} 1530 nm, \textbf{g,} 1532 nm, \textbf{h,} 1556 nm, and \textbf{i,} reference at 1570 nm.}
\label{fig4}
\end{figure}

Leveraging the dispersion engineering capability of the a-SiC/LN platform, it is possible to substantially enlarge the group index of the waveguide mode near the PhC band edge. Consequently, the reduction in group velocity enhances the effective electro-optic light-matter interaction per unit waveguide length, leading to greater phase accumulation and more efficient EO tuning.\cite{vlasov2005active,li2025ultra,baba2008slow} As shown in Fig. 4\textbf{a} and \textbf{b}, a push-and-pull configured MZI, with its two tuning arms embedded with the same PhC sections (leakage-uniform-leakage) as discussed in the previous section, is employed to correlate the slow-light effect with the EO tuning efficiency. On the fabricated slow-light MZI (Fig. 4\textbf{b}) with a PhC length of $L_\text{PhC}=$ 1.013 mm and an electrodes gap of $g_\text{2}$ = 4.0 $\mu m$, the passive transmission spectrum is measured before the EO measurements, as shown in Fig. 4\textbf{c}. A photonic bandgap with an extinction ratio of roughly 25 dB is obtained over the wavelength range of 1535 - 1553 nm. 

The electro-optic tuning efficiencies are then measured by applying a voltage difference ranging from 0 to 60 V across the MZI electrodes at different wavelengths. The measured half-wave voltage-length product results are shown in Fig. 4\textbf{d}, together with the corresponding effective group index values extracted based on Eq.(3), which are shown in Fig. 4\textbf{e}. 
Near the band edge around 1535 nm, half-wave voltage–length products of 2.30 V$\cdot$cm at 1530 nm and 1.38 V$\cdot$cm at 1532 nm are determined, corresponding to group indices of 12.91 and 21.34 respectively, as shown in Fig. 4\textbf{f} and \textbf{g}. Near the other band edge around 1553 nm, we measure half-wave voltage–length products of 3.97 V$\cdot$cm at 1556 nm, as shown in Fig. 4\textbf{h}, which corresponds to a group index of 7.59. 
Inside the passband, at 1570 nm a half-wave voltage-length product of 11.84 V$\cdot$cm is measured and a group index of 2.57 is extracted (Fig. 4\textbf{i}), as a reference EO tuning efficiency of the MZI device without slow-light enhancement. 
The detailed parameters of the slow-light MZI and more measurement results at different wavelengths are included in Section V of the Supplementary Materials. 
Compared to the reference working point at 1570 nm, the group index enhancement is enhanced by a factor of approximately 8.5, yielding an improvement in electro-optic tuning efficiency by the same factor. The reduction in the EO half-wave voltage-length product values provides experimental evidence of the presence of the slow-light effect. Further reduction of the half-wave voltage-length product is expected to be achieved by increasing the number of slow-light PhC hole periods embedded in the MZI tuning arms.\cite{li2025ultra} Nevertheless, the experimental results demonstrate that partially etched a-SiC/LN PhC structures enable effective control of optical confinement and exceptional dispersion-engineering capability. Excellent fabrication consistency and long-term process stability are verified by comparing two devices fabricated in two separate batches over a time span of six months (Supplementary, Section VI), showing the robustness and stability of the optimized a-SiC/LN fabrication processes. A comparison between this monolithic a-SiC/bulk-LN platform with other existing high-density LiNbO$_3$-based PIC platforms is shown in Table.S3 in Supplementary. These  results highlight the strong potential of the monolithic a-SiC/LN platform as a scalable and cost-effective foundation for next-generation photonic technologies.

\section{Discussion and conclusion}\label{secconclusion}

In this work, we demonstrate the fully monolithic a-SiC/LiNbO$_3$ platform, enabling high-density, efficient EO tunable, LiNbO$_3$-etching-free, CMOS compatible and scalable photonic integration directly on bulk LiNbO$_3$ crystals. The fabricated devices demonstrate competitive performances in waveguide propagation losses, thermo-optic and electro-optic tuning efficiencies, while partially etched a-SiC/LiNbO$_3$ photonic crystal structures further extend the platform capability to optical-confinement control and dispersion engineering. By eliminating the reliance on ion-sliced thin-film LiNbO$_3$ and wafer-bonding processes, this fully monolithic a-SiC/LN platform substantially reduces fabrication complexity and cost, clearing key barriers toward LiNbO$_3$ PIC's rapid scale-up. Beyond bulk LiNbO$_3$, the strategy of monolithically integrating a-SiC rib-loading waveguides on low-index materials can be extended to other functional photonic substrates that face similar challenges in thin-film formation and etching, providing a CMOS-compatible and scalable route toward high-performance photonic integrated circuits on a broad class of material platforms.

\section*{Methods}\label{secmethods}

\subsection*{Devices fabrication}
We prepared the LN substrates by dicing a 4-inch bulk lithium niobate wafer (Biotain Crystal) into samples with sizes of 14 $\times$ 12 mm$^2$. After dicing, the samples were cleaned in the standard Piranha solution at 60 $^\circ C$. On the diced samples, we deposited 350 nm a-SiC via ICPCVD (Oxford PlasmaPro 100) at 150 $^\circ C$. 
To fabricate the photonic devices on the a-SiC layer, electron-beam (E-beam) resist ARP6200-13 was spin-coated, followed by sputtering a 5 nm chromium (Cr) layer on top of the resist, as shown in fig. 1\textbf{a}. This Cr layer serves as a charge dissipator layer during electron-beam lithography (EBL). Together with proximity correction, it guarantees high-resolution and high-fidelity pattern definition. We performed E-beam lithography (Raith 100-kV EBPG-5200), after which the Cr passivation layer was removed using a solution of perchloric acid and ceric ammonium nitrate. Then, the exposed resist was developed using Pentyl Acetate, MIBK:IPA = 1:1, and IPA. The EBL defined PhC holes were transferred to the a-SiC layer by reactive-ion etching (RIE, Sentech Etchlab 200), using a gas mixture of SF$_6$ and O$_2$.\cite{lopez2023high,li2025heterogeneous} The a-SiC waveguides were fabricated by repeating the same process sequence (spin-coating, Cr sputtering, EBL exposure, Cr removal, development, dry etching). 
Afterward, another EBL step and metal liftoff were carried out to define the electrodes. The electrodes consist of a 5 nm Cr adhesion layer and a 350 nm gold layer. Finally, a 1.2 $\mu$m layer of SiO$_2$ cladding is deposited by ICPCVD. Openings in the cladding are defined by lift-off process, leaving the electrode pads exposed for wire-bonding, as previously described in \cite{li2025heterogeneous}.

\subsection*{Characterization}
The optical responses of the a-SiC/LN devices are characterized using the setup depicted in Fig. 2\textbf{b}. Telecom light is coupled from tunable laser (Photonetics TUNICS-PRI 3642), through single-mode fiber (SMF-28) and a fiber polarization controller (FPC 560), to the device under test (DUT) via the on-chip apodized grating couplers. The DUT is mounted and fixed on a printed circuit board, with electronic connections to a direct-current voltage source (RIGOL DP832A) and a proportional-integral-derivative temperature controller. For thermo-optical measurements, silver paste is used to ensure good thermal conductivity between the DUT and the PCB. For electro-optical measurements, the on-chip electrodes are wire-bonded with the metal pads on the PCB. The output light from the DUT is coupled via a second grating coupler into a multi-mode fiber to maximize collection efficiency, before being detected by an optical powermeter (Newport 818-NR).

\subsection*{PhC mirror reflection extraction}

The reflectivity of the PhC mirrors is extracted from the Fabry-P\'{e}rot resonance fitting (F-P fit) of the measured spectra shown in Fig. 3\textbf{f}, assuming that the two PhC mirrors forming the Fabry-P\'{e}rot cavity hold identical reflectivity. The transmission $T$ at the through port of the device depicted in Fig. 3\textbf{d} can be written as \cite{rabus2020integrated}: 
\begin{equation}
T = k\lambda+T_0+\left|t - \frac{\kappa^2tR^2\cdot e^{-\alpha L}\cdot e^{i4\pi n_{\text{eff}}L/\lambda}}{1-t^2R^2\cdot e^{-\alpha L}\cdot e^{i4\pi n_{\text{eff}}L/\lambda}}\right| ^2, 
\label{eq1}
\end{equation}
where $t$ and $\kappa$ represent the direct transmission factor and the coupling factor at the evanescent coupling region, satisfying $t^2 + \kappa^2 = 1$. The distance between the two PhC mirrors is described by $L$, while $\alpha$ corresponds to the power attenuation coefficient of the propagating mode. To extract the one-side PhC mirror reflectivity $R$, the fundamental TE mode effective index $n_{\text{eff}}$ is substituted into the equation. A linear term $kx+b_1$ is included in the F-P fit model to describe the linear background slope that arises from intracavity interferences and grating coupler responses. 

To extract the cavity quality factor, the F-P resonances are fitted with a Lorentzian function in the form of: 
\begin{equation}
T = T_0+m\lambda-\frac{A}{1+(\frac{2(\lambda-\lambda_0)}{FWHM})^2},
\label{eq2}
\end{equation}
where $T_0+m\lambda$ describes the linear background in the measured spectra, $\lambda_0$ represents the resonant wavelength, and $A=1-T_\text{min}$ denotes the transmission extinction at resonance. $FWHM$ is the full-width half-maximum of the resonance dips extracted for quality factor calculation. The same Lorentzian function is also used for extracting quality factors of the ring resonators.

\subsection*{Effective group index and slow-light}
The optical phase difference between two tuning arms of an MZI stems from the linear Pockels effect, namely $\Delta\phi = \frac{2\pi}{\lambda}\Delta n_\text{eff} L$, where $L$ is the tuning length. In an a-SiC on X-cut LN slow-light waveguide, the effective index change of the propagating mode can be expressed by $\Delta n_\text{eff} = \Gamma \cdot\frac{1}{2} n_g n_e^2r_{33}\frac{V}{g}$, where $\Gamma$ is the optical-electrical field overlap factor, and $g$ is the electrode gap. The group index change, shown in Fig.~\ref{fig4}e, therefore can be extrapolated from the measured half-wave voltage-length product of the electro-optical tuning (Fig. 4\textbf{d}), based on the equation\cite{li2025ultra}: 
\begin{equation}
n_g = \frac{\lambda \cdot g}{2\cdot \Gamma \cdot V_\pi L \cdot n_e^2 \cdot r_{33} }.
\label{eq3}
\end{equation}
Here, $V_\pi L$ is the EO tuning efficiency of a push-and-pull configured MZI.

\backmatter

\section*{Supplementary information}

Supporting information can be found in Supplementary Information.

\section*{Data availability}

The data supporting the findings of this study are publicly available in the Zenodo repository at https://doi.org/10.5281/zenodo.21792757.

\section*{Acknowledgements}

The authors acknowledge the support provided by the staff of the Kavli Nanolab Delft. Z.L. acknowledges the China Scholarship Council (CSC, 202206460012). I. E. Z. acknowledges funding from the European Union’s Horizon Europe research and the innovation programme under grant agreement No. 101098717 (RESPITE) and 101099291 (fastMOT).

\section*{Author contributions}

Z.L. and I.E. conceived the idea. Z.L. and H.S. conducted the design and simulations. Z.L. fabricated the devices. Z.L. performed the measurements. B.L. and T.S. contributed to the building of the experimental setups. I.E. and S.G. supervised the project. Z.L. wrote the original manuscript. All authors participated in reviewing and editing the manuscript.

\section*{Competing interests}

The authors declare no competing interests.

\bibliography{sn-bibliography}

\end{document}